\documentclass[12pt]{article}

\usepackage{color}
\usepackage{amsmath}
\usepackage{amssymb}
\usepackage{graphicx}
\usepackage{dcolumn}
\usepackage{bm}
\usepackage{hyperref}
\usepackage{nameref}
\usepackage{multirow,ulem}
\usepackage{array}
\usepackage{ulem}
\usepackage{authblk}
\usepackage{nameref}
\usepackage{setspace}
\usepackage[margin=1in]{geometry}

\newcolumntype{P}[1]{>{\centering\arraybackslash}p{#1}}

\title{Response to an external field of a generalized Langevin equation with stochastic resetting of the memory kernel}

\author{Petar Jolakoski\textsuperscript{1}, Lasko Basnarkov\textsuperscript{2}, Ljupco Kocarev\textsuperscript{1,2}, Aleksandra Popovska-Mitrovikj\textsuperscript{2}, Verica Bakeva\textsuperscript{2}, Trifce Sandev\textsuperscript{1,3,4}}

\affil{\footnotesize
\textsuperscript{1}\footnotesize Research Center for Computer Science and Information Technologies, Macedonian Academy of Sciences and Arts, Bul. Krste Misirkov 2, 1000 Skopje, Macedonia \\
\textsuperscript{2}\footnotesize Faculty of Computer Science and Engineering, Ss. Cyril and Methodius University, PO Box 393, 1000 Skopje, Macedonia \\
\textsuperscript{3}\footnotesize Institute of Physics, Faculty of Natural Sciences and Mathematics, Ss.~Cyril and Methodius University in Skopje, Arhimedova 3, 1000 Skopje, Macedonia \\
\textsuperscript{4}\footnotesize Department of Physics, Korea University, Seoul 02841, Korea}

\date{April 9, 2025}

\begin{document}

\newtheorem{theorem}{Theorem}
\newtheorem{definition}{Definition}
\newtheorem{lemma}{Lemma}
\newtheorem{proposition}{Proposition}
\newtheorem{remark}{Remark}
\newtheorem{corollary}{Corollary}
\newtheorem{example}{Example}

\maketitle
\begin{abstract}

We study a generalized Langevin equation (GLE) framework that incorporates stochastic resetting of a truncation power-law memory kernel. The inclusion of stochastic resetting enables the emergence of resonance phenomena even in parameter regimes where conventional settings (without resetting) do not exhibit such behavior. Specifically, we explore the response of the system to an external field under three scenarios: (i) a free particle, (ii) a particle in a harmonic potential, and (iii) the effect of truncation in the memory kernel. In each case, the primary focus is on understanding how the resetting mechanism interacts with standard parameters to induce stochastic resonance. In addition, we explore the effect of resetting on the dielectric loss. 

\end{abstract}

\maketitle

\section{Introduction}\label{sec:GLE}

The random motion of a massive particle of mass $m$ in a fluid, known as Brownian motion, can be described using the Langevin equation~\cite{langevin1908theorie}. It is an equation that combines Newton's second law with the random force $\zeta(t)$, which is a white Gaussian noise of zero mean\footnote{The notation $\langle\cdot\rangle$ denotes an ensemble average, which is a statistical averaging over an ensemble of particles at a given moment in time $t$.}, $\langle\zeta(t)\rangle=0$, and the delta correlation, $\langle\zeta(t)\zeta(t^{\prime})\rangle=2k_BT\gamma\delta(t-t^{\prime})$, where $k_B$ is the Boltzmann constant and $T$ is the absolute temperature of the environment. Thus, the standard Langevin equation reads
\begin{align}
    m\ddot{x}(t)+\gamma \dot{x}(t)=\zeta(t)
\end{align}
where $-\gamma\dot{x}(t)$ is the friction force, which is the result of the interaction between the particle and the surrounding molecules of the environment, and $\gamma$ is the friction coefficient. The mean squared displacement (MSD), $\langle x^2(t)\rangle$ of the Brownian particle has a linear dependence on time. This is a simple random process with normal diffusive behavior. However, in many complex media, the MSD shows a power-law dependence on time, $\langle x^2(t)\rangle\sim t^{\alpha}$, which is a signature of anomalous diffusion. Here, $\alpha$ is the anomalous diffusion exponent, which distinguishes the cases of subdiffusion for $0<\alpha<1$, normal diffusion for $\alpha=1$ and superdiffusion for $\alpha>1$~\cite{metzler2000random}. There are many different approaches to anomalous diffusion, such as the continuous time random walk (CTRW) model~\cite{metzler2000random,barkai2000continuous}, fractional diffusion and Fokker-Planck equations~\cite{metzler2000random,metzler1999anomalous,barkai2000continuous}, fractional Brownian motion~\cite{kolmogorov1940wienersche,mandelbrot1968fractional}, fractional and generalized Langevin equations~\cite{lutz2001fractional,deng2009ergodic}, to name a few. 

Here, we consider the GLE for a particle with mass $m$ in external potential $V(x)$, which is defined by
\begin{equation}\label{GLEgeneral}
    m\ddot{x}(t) + \int_0^t \gamma(t - t^{\prime}) \dot{x}(t^{\prime}) \, dt^{\prime} + \frac{dV(x)}{dx} = \zeta(t), \quad \dot{x}(t) = v(t),
\end{equation}
where $x(t)$ is the particle displacement, $v(t)$ is the particle velocity, $F(x)=-\frac{dV(x)}{dx}$ is the external force acting on the particle, $\gamma(t)$ is the friction memory kernel and $\zeta(t)$ is a random force of zero mean $(\langle \zeta(t) \rangle= 0)$ which satisfies the second fluctuation-dissipation relation \cite{kubo2012statistical,zwanzig2001nonequilibrium}
\begin{equation}\label{FDR}
\langle \zeta(t)\zeta(t^{\prime}) \rangle = C (|t - t^{\prime}|) = k_B T\gamma{(|t-t^{\prime}|)}.
\end{equation}
This fluctuation-dissipation relation means that the dissipation represented by the friction memory kernel, and the fluctuation, represented by the noise $\zeta(t)$, come from the same source\footnote{If the second fluctuation-dissipation relation~(\ref{FDR}) does not hold then the noise is known as external noise.}. The GLE extends the standard Langevin equation by introducing a memory-dependent friction term, thus accounting for non-Markovian effects in stochastic processes. The friction kernel reflects how a complex environment affects the motion of the particles, resulting in anomalous diffusion. Because of this, the GLE provides a useful framework for modeling such diffusion in various complex systems. Essentially, it represents Newton's second law for a particle under random force and generalized friction. 

The memory kernel in the GLE may have different forms. For a Dirac $\delta$ memory kernel, the GLE reduces to the standard Langevin equation for a Brownian motion. This $\delta$ correlation of the noise occurs when the mass of the immersed particle (for example, pollen grains and dust particles) in a fluid (for example, water) is much larger than the mass of the surrounding molecules. The time scale of the molecular motion is much shorter than the time scale of the Brownian motion. However, in many systems the mass of the immersed particle is not necessarily much larger than the mass of the surrounding molecules. Thus, the time scale of the molecular motion is not
very much shorter than the time scale of the motion of the immersed particle, and the noise correlation is no longer $\delta$ correlated. The power-law correlated noise $\gamma(t)=\gamma\frac{t^{-\alpha}}{\Gamma(1-\alpha)}$, $0<\alpha<1$, is an example which leads to anomalous diffusive behavior. In such a case the free particle shows anomalous diffusive behavior since the mean squared displacement (MSD) has a power-law dependence on time, i.e., $\langle x^2(t)\rangle\sim t^{\alpha}$~\cite{lutz2001fractional}, see also Refs.~\cite{vinales2006anomalous,desposito2008memory,burov2008fractional,burov2008critical}.

The GLE has been used to model anomalous dynamics within a single protein molecule~\cite{kou2004generalized,10.1214/07-AOAS149,min2005observation}, in which a power-law memory kernel was used, and a harmonic potential $V(x)=\frac{m\omega^2x^2}{2}$, where $\omega$ is the oscillator frequency, since the movement within proteins is confined to a short range which can be well approximated by a harmonic potential. It can also be used to describe the single-file diffusion in which the MSD behaves as $\langle x^2(t)\rangle\sim t^{1/2}$, the motion in a viscoelastic environment~\cite{goychuk2009viscoelastic,goychuk2012viscoelastic}, the individual motion of lipid molecules~\cite{kneller2011communication,jeon2011vivo,jeon2012anomalous} and of messenger RNA molecules~\cite{weber2010bacterial} in living cells, etc. 

In recent years, a very popular topic in non-equilibrium statistical physics is the influence of stochastic resetting on the corresponding random process. That is, a process where the moving particle is reset to its initial position or any other position from time to time. Such a resetting mechanism can make the mean first passage time of a random walker, for example a Brownian walker, to hit the target finite, which diverges in the reset-free case~\cite{evans2011diffusion}. Stochastic resetting also leads the particle to reach a non-equilibrium stationary state in the long time limit~\cite{evans2011diffusion,evans2020stochastic}, which has also been experimentally demonstrated by using holographic optical
tweezers~\cite{tal2020experimental} and laser traps~\cite{besga2020optimal}. Relaxation to the non-equilibrium stationary state appears to be a non-trivial process, showing a dynamical phase transition~\cite{majumdar2015dynamical}. Such phenomena have also been observed in various inhomogeneous and disordered systems, such as heterogeneous diffusion processes~\cite{lenzi2022transient,sandev2022heterogeneous,ray2022expediting,pal2024random}, geometric Brownian motion~\cite{stojkoski2021geometric,vinod2022nonergodicity,vinod2022time}, random walks in complex networks~\cite{riascos2020random,huang2021random,guerrero2025random,michelitsch2025random}, and in quantum systems~\cite{mukherjee2018quantum,yin2023restart,kulkarni2023generating}, to name a few. However, there are a limited number of works on the GLE with resetting. In the recent paper~\cite{biswas2025resetting}, the GLE for a particle in a viscoelastic bath under stochastic resetting was considered, where the memory kernel is a combination of the Dirac delta and exponential function. In this paper, we consider a GLE in which we use the renewal equation for the memory kernel, as it is done in Ref.~\cite{petreska2022tuning} to tune the dielectric relaxation and complex susceptibility in a system of polar molecules.    

This paper is organized as follows. In Section~\ref{sec2}, we review the known results for the GLE for a harmonic oscillator in the presence of an external periodic force. We investigate the stochastic resonance for a tempered power-law memory kernel by calculating the response function and the complex susceptibility. In Section \ref{sec:gle-reset}, we provide a detailed description of how stochastic resetting of the memory kernel is incorporated into the generalized Langevin equation. We analyze the effects of resetting on stochastic resonance and the dielectric loss, focusing on distinct cases: free particle, harmonically bounded particle, truncation and their interplay with the resetting rate of the memory kernel. The double-peak phenomena were observed. The analytical findings are verified through the Markovian embedding simulation methodology scheme developed in Section \ref{sec:simulation-methodology}. Finally, in Section \ref{sec:discussion}, we summarize our findings and propose potential avenues for future research.

\section{GLE for a harmonic oscillator: Response to an external periodic force}\label{sec2}

The GLE for a harmonically bounded particle with mass $m=1$ and an external periodic force is given by~\cite{burov2008critical,burov2008fractional}:
\begin{equation}\label{eq:gle-free}
    \ddot{x}(t) + \int_0^t \gamma(t - t^{\prime}) \dot{x}(t^{\prime}) \, dt^{\prime} + \omega^2 x(t) = A_0 \cos(\Omega t) + \zeta(t), \quad \dot{x}(t) = v(t).
\end{equation}

The term $\omega^2 x(t)$ originates from the harmonic potential $V(x)=\frac{m\omega^2x^2}{2}$, which produces a restoring force proportional to the particle displacement. Setting the total external force $F(x,t)=-\omega^2x(t)+A_0\cos(\Omega t)$ equal to zero, we obtain the standard GLE equation~(\ref{GLEgeneral}). 
Here, we consider exponentially truncated power-law memory kernel of the form\footnote{A graphical representation of the truncated memory kernel, together with resetting, is given in Fig.~\ref{fig:kernels-visualized}, which is discussed in greater detail in Sec~\ref{sec:gle-reset}.}:
\begin{equation} \label{eq:pl-kernel}
    \gamma(t) = \gamma e^{-bt}\frac{t^{-\alpha}}{\Gamma (1-\alpha)},
\end{equation}
which for $b=0$ reduces to the power-law memory kernel. The effect of truncation is captured by the truncation parameter $b$, whereas the subdiffusive behavior is controlled by the power-law exponent $\alpha$. Its Laplace transform reads
\begin{equation} \label{eq:pl-kernel-laplace}
    \hat{\gamma}(s) = (s+b)^{\alpha-1},
\end{equation}
where we apply the shift rule of the Laplace transform,
$\mathcal{L} \left[e^{-at} f(t) \right] = \hat{f}(s + a)$,
where $\hat{f}(s) = \mathcal{L}[f(t)]=\int_{0}^{\infty}f(t)e^{-st}dt$. Memory kernels, such as Eq.~\eqref{eq:pl-kernel} and more generalized truncated power-law forms have been used in the generalized diffusion equation~\cite{sandev2015diffusion,sandev2017beyond,petreska2022tuning} and generalized Langevin equation~\cite{liemert2017generalized,sandev2017generalized,molina2018crossover} for description of characteristic crossover dynamics.

In general, a system subjected to a time-dependent oscillating field tends to exhibit stochastic resonance \cite{gammaitoni1998stochastic}, when the frequency of the external field coincides with one of the system's intrinsic natural frequencies. For a standard power-law memory kernel, when $\alpha=1$ resonance occurs only if the angular frequency $\omega$ of the potential exceeds the critical value $\gamma / \sqrt{2}$. However, in the case of fractional dynamics when $0 < \alpha < 1$ there is a non-trivial behavior of the resonance phenomenon. Specifically, in \cite{burov2008fractional}, the authors identified a critical exponent $\alpha_R$ such that resonance occurs when $\alpha<\alpha_R$. Interestingly, the resonant peak is present even in the case of a free particle (absence of a potential). On the other hand, for $b>0$ resonance dissapears as the truncation parameter $b$ becomes larger \cite{liemert2017generalized}. Moreover, additional critical exponents are determined for the onset of the double peak phenomenon in the imaginary part of the complex susceptibility. This quantity is examined as a signature of magnetic (or dielectric) loss arising from thermally activated relaxation processes in magnetic nanoparticles—whether the anisotropy is uniaxial, cubic, or triaxial—and under various external field conditions \cite{kalmykov2003analytic,kalmykov2020dipole, kalmykov2016dynamic,kalmykov2005longitudinal,kalmykov1998longitudinal}.

In this paper, we examine how stochastic resetting of the memory kernel affects the response of the system described by Eq.~\ref{eq:gle-free} to an external oscillating force. Stochastic resetting is a mechanism in which a given stochastic process evolves freely during a given random interval of time before being reset to its initial state \cite{evans2011diffusion,evans2020stochastic}. It has motivated widespread research in statistical physics and stochastic processes. One key application is to enforce stationarity in processes that are inherently non-stationary \cite{stojkoski2021geometric,jolakoski2025impact}, and another is that introducing resetting causes the mean first passage time (MFPT) for a single diffusive searcher to become finite - in contrast with purely diffusive searches where the MFPT diverges \cite{evans2011diffusion,evans2011optimal}. In our case, we apply resetting directly to the memory kernel instead of to the particle's position, as is done in \cite{petreska2022tuning} in investigation of dielectric relaxation dynamics, and explore its consequences on the stochastic resonance phenomenon and the dielectric loss. We find that the resetting rate, by influencing the correlation structure of the noise through the memory kernel, alters the dynamical transitions from a non-resonance to a resonance regime. Particularly, we observe the appearance of resonance-like behavior for regions where it was previously absent ($\alpha>\alpha_R$). Similarly, the resetting mechanism interacts with the angular frequency of the potential in an interesting way, inducing resonance that was previously absent for higher values of the truncation parameter $b$. Additionally, we identify critical regions in the $\alpha-r$ plane for the imaginary part of the complex susceptibility, in relation to other standard parameters of the GLE.

Here, we briefly review some of the known results of a response to an external field and complex susceptibility in the standard GLE. 
The starting point for this analysis is to perform an average of Eq.~\ref{eq:gle-free} and using the Laplace transformation method, for the first moment one finds:
\begin{equation}
\langle x(t) \rangle = x_0 \left[ 1 - \omega^2 \int_0^t h(t^{\prime}) \, dt^{\prime} \right] + v_0 \, h(t) + A_0 \int_0^t \cos \left( \Omega (t - t^{\prime}) \right) h(t^{\prime}) \, dt^{\prime},
\end{equation}
where \( h(t) = \mathcal{L}^{-1} \left[ \hat{h}(s) \right] = \mathcal{L}^{-1} \left[ \frac{1}{s^2 + s \, \hat{\gamma}(s) + \omega^2} \right] \). From here, for the long time limit (\( s \to 0 \)) it follows:
\begin{equation}\label{eq:lt-response}
\langle x(t) \rangle \simeq A_0 \int_0^t \cos \left( \Omega (t - t^{\prime}) \right) h(t^{\prime}) \, dt^{\prime} \quad \rightarrow \quad \langle x(t) \rangle = R(\Omega) \cos \left( \Omega t + \theta(\Omega) \right), \quad t \to \infty,
\end{equation}
where the response $R(\Omega)$ and the phase shift $\theta(\Omega)$ are defined through the complex susceptibility:
\begin{equation} \label{eq:complex-susceptibility}
    \chi(\Omega) = \chi^{\prime}(\Omega) + i\chi^{\prime \prime}(\Omega) = \hat{h}(-i\Omega),
\end{equation}
where $\chi^{\prime}(\Omega)$ and $\chi^{\prime \prime}(\Omega)$ are the complex and imaginary parts of the susceptibility, and $\hat{h}(-i\Omega) = \int_0^{\infty} e^{i\Omega t}h(t)$. From the complex susceptibility, the response is 
\begin{align}
    R(\Omega)=| \chi(\Omega)|
\end{align} 
and the phase shift is 
\begin{align}
    \theta(\Omega) = \arctan\left(-\frac{\chi^{\prime \prime}(\Omega)}{\chi^{\prime}(\Omega)}\right).
\end{align}
The response and the real and imaginary parts of the complex susceptibility will be studied in detail.

Generally, extensive research has focused on understanding how subdiffusing systems respond to a time-dependent field \cite{barbi2005linear,sokolov2006field,heinsalu2007use}. For our particular focus, the authors in \cite{burov2008fractional,burov2008critical,liemert2017generalized} focus on the response of GLE with and without harmonic potential and also investigate the behavior of the imaginary part of the complex susceptibility (the dielectric loss). To calculate the complex susceptibility for the standard case of GLE with truncated power-law memory kernel we substitute the Laplace transform of the memory kernel (Eq.~\ref{eq:pl-kernel-laplace}) in $\hat{h}(s)$, that is $\hat{h}(s) = \frac{1}{s^2 + s \, \hat{\gamma}(s) + \omega^2}$, and calculate the complex susceptibility as defined by Eq.~\ref{eq:complex-susceptibility}. The result of this calculation is
\begin{align}\label{eq:standard-case-susceptibility}
    \chi(\Omega) = \hat{h}(s=-i\Omega) = \frac{1}{(-i\Omega)(-i\Omega+b)^{\alpha-1}-\Omega^2 + \omega^2}.
\end{align}

\begin{figure}[h!]
\centering
\includegraphics[width=12cm]{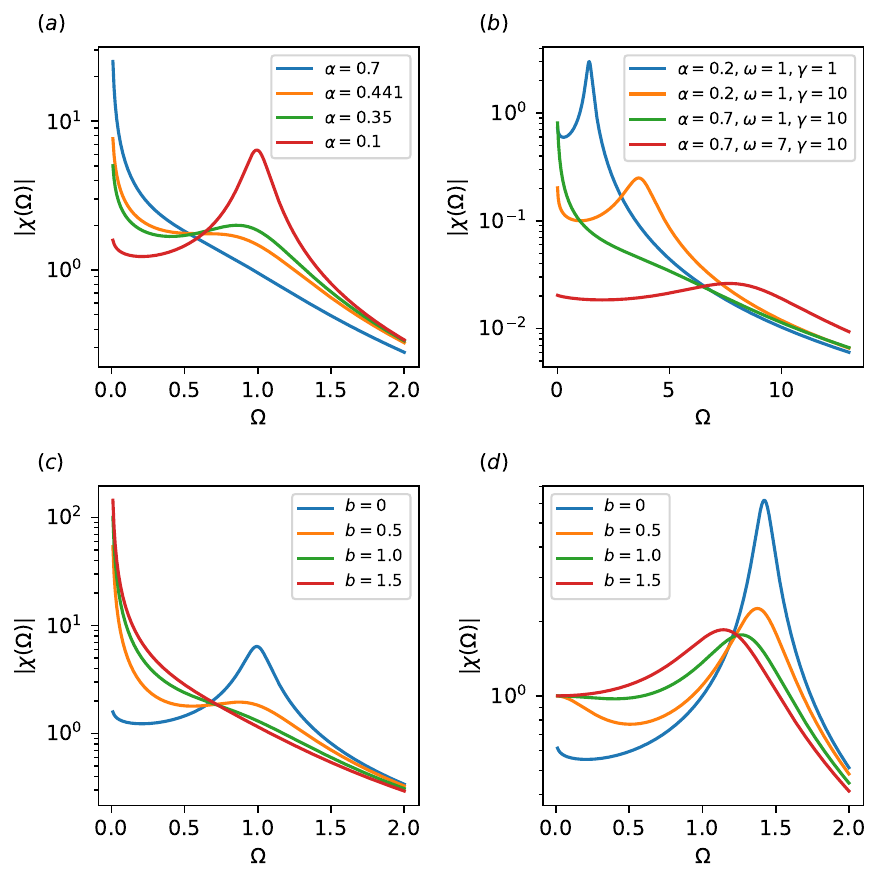}
\caption{The response of GLE to an oscillating field. Subplot $(a)$ shows the response for a free particle ($\omega=0$). On the other hand, $(b)$ shows the effect for a harmonically bounded particle for various parameters. $(c)$ shows the effect of the truncation parameter for a free particle. $(d)$ incorporates the effect of the potential $(\omega=1)$ for various truncation parameters. $\alpha=0.1$ for both $(c)$ and $(d)$.}
\label{fig:sr-standard-case}
\end{figure}

The case without truncation ($b=0$) was analyzed in great detail in \cite{burov2008fractional,burov2008critical}. On the other hand, the effect of truncation ($b>0$) on the susceptibility was considered in \cite{liemert2017generalized}. First, as described in Sec.~\ref{sec:GLE}, the response is calculated from the complex susceptibility as $R(\Omega)=|\chi(\Omega)|$. This quantity is used to detect the presence of stochastic resonance. Particularly, for the normal diffusion case (when $\alpha=1$) the response is a decaying function of the frequency of the external force, $\Omega$. However, for $0 < \alpha < 1$, the response $R(\Omega)$ can obtain a maximum for some positive $\Omega_R>0$ which means that the system exhibits resonance at that frequency $\Omega_R$. Interestingly, this happens also for the case of a free particle ($\omega=0$) i.e. in the absence of harmonic potential. The response, calculated as the magnitude of Eq.~\ref{eq:standard-case-susceptibility}, is

\begin{align} \label{eq:response-standard-case}
|\chi(\Omega)| 
=
\frac{1}{
\sqrt{\Bigl[\omega^{2} - \Omega^{2} \;+\;\Omega\,z^{\alpha-1}\,\cos(\theta)\Bigr]^{2}
\;+\;\Bigl[\Omega\,z^{\alpha-1}\,\sin(\theta)\Bigr]^{2}}},
\end{align}
where $z = \sqrt{\,b^{2} + \Omega^{2}}\,$ and $\displaystyle \theta = (\alpha -1)\,\arctan \Bigl(\frac{-\,\Omega}{b}\Bigr) - \tfrac{\pi}{2}\,$. Taking the limit of Eq.~\ref{eq:response-standard-case} as $b\to 0$, we recover the response calculated in~\cite{burov2008fractional}. For free motion and $b=0$, it was found that a resonance exists if $\alpha<\alpha_R=0.441021\dots$, as shown in Fig.~\ref{fig:sr-standard-case}(a). It was also argued that this finding is consistent with the interpretation that, for small fractional exponents $\alpha$, the friction force can be viewed as an effective elastic force arising from the ``cage'' effect~\cite{burov2008fractional}. In addition, points of maximum of $R(\Omega)$ generally depend on $\omega$ and $\gamma$ (Fig.~\ref{fig:sr-standard-case}(b)). Concretely, the presence of a harmonic potential creates a favorable environment for resonance to occur, even in the case $\alpha>0.441$. However, as the friction constant $\gamma$ increases, resonance may be disrupted and the response is reduced. Similarly, if the truncation parameter becomes active $b>0$, then the amplitude of the response becomes smaller and resonance slowly disappears (Fig.~\ref{fig:sr-standard-case}(c)) for a free particle. The potential can reverse this effect by generating conditions that promote resonance (Fig.~\ref{fig:sr-standard-case}(d)).

After briefly reviewing the results for the standard case, we continue with the main analysis of this paper: exploring how resetting of the memory kernel influences stochastic resonance and its interplay with other standard parameters.

\section{Stochastic resonance under resetting of the memory kernel} \label{sec:gle-reset}

In this paper, we extend the model by incorporating an exponential resetting of the memory kernel, as is done in~\cite{petreska2022tuning}. Under these resetting dynamics, the particle follows a GLE (Eq.~\ref{eq:gle-free}), but the memory kernel is intermittently reset to its initial configuration, therefore erasing any accumulated history-dependent influences. The resetting of the memory kernel can be interpreted as removing any modifications or evolution that the kernel may have undergone due to interactions, external perturbations, or changes in the system's environment since its initial state. Mathematically, this is done as follows:
\begin{equation} \label{eq:kernel-resetting}
    \gamma_r(t) = e^{-rt}\gamma(t) + \int_0^t r e^{-rt^{\prime}} \gamma(t^{\prime}) \,dt^{\prime}
\end{equation}
where $r$ is the resetting rate. This means that the resetting time is sampled from an exponential (Poissonian) distribution $p(t^{\prime}) = r e^{-rt^{\prime}}$, with resetting rate $r > 0$. Its Laplace transform reads:
\begin{equation}
    \hat{\gamma}_r(s) = \frac{s+r}{s} \hat{\gamma}(s+r),
\end{equation}
where we apply the shift rule of the Laplace transform. From eqs.~\ref{eq:pl-kernel} and \ref{eq:pl-kernel-laplace} it follows that:
\begin{equation}\label{eq:pl-kernel-laplace-reset}
    \hat{\gamma}_r(s) = \frac{s+r}{s} (s+r+b)^{\alpha-1}
\end{equation}
Eq.~\ref{eq:pl-kernel-laplace-reset} will play a crucial role in the main calculations presented later in this paper.

\begin{figure}[h!]
\centering
\includegraphics[width=17cm]{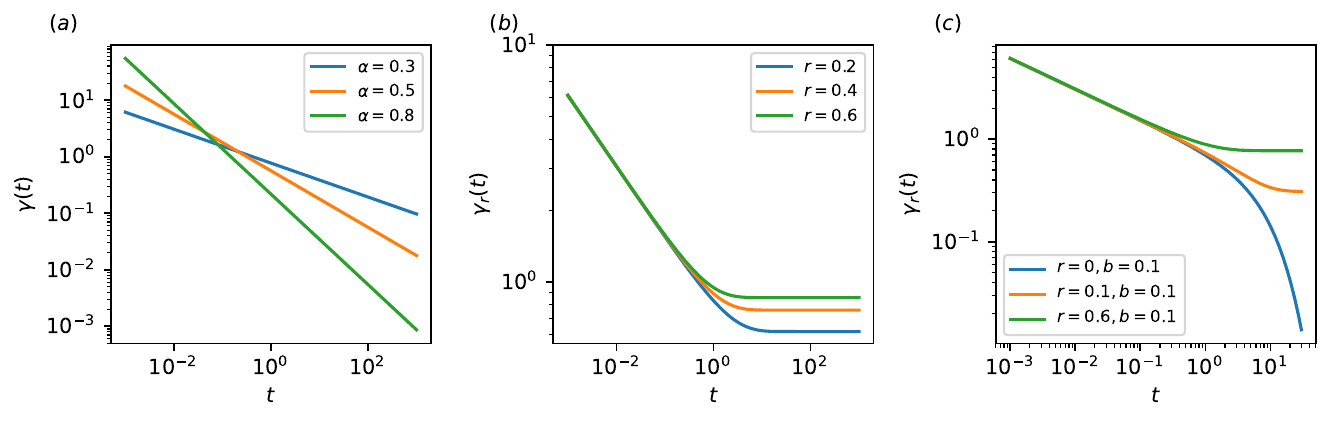}
\caption{Log-log plot of the memory kernel Eq.~\ref{eq:kernel-resetting} for (a) without resetting, (b) with resetting for $\alpha=0.3$ and (c) with truncation and various resetting rates for $\alpha=0.3$.}
\label{fig:kernels-visualized}
\end{figure}

Before proceeding to the main analysis of the paper, we briefly review the graphical representations of the memory kernel under various parameter settings. In the case without resetting Fig.~\ref{fig:kernels-visualized}(a), as $\alpha$ increases the slope of the kernel becomes steeper, causing $\gamma(t)$ to decay more rapidly. Conversely, smaller $\alpha$ values lead to slower decay in the intermediate-time regime, indicating stronger long-memory effects. Once resetting is introduced, Fig.~\ref{fig:kernels-visualized}(b), the parameter $r$ damps out the long-time tail of the kernels. The kernel saturates at a nonzero value, which implies that there is a constant memory that does not vanish at long times. Finally, in Fig.~\ref{fig:kernels-visualized}(c), a truncation is introduced by setting $b$ to 0.1. Compared to the purely power-law decay (when $b=0$), one observes an exponential cutoff at longer times. As $r$ increases, the memory kernel retains its larger amplitude for timescales up to $1/b$, before ultimately decaying due to the exponential factor $e^{-bt}$. This interplay between power-law and exponential behavior reflects the competition between fractional memory effects and exponential relaxation.

We extend our analysis to the problem of stochastic resonance with resetting. In particular, we consider exponential resetting of the memory kernel (Eq.~\ref{eq:kernel-resetting}). Following the approach used earlier for the standard case in Sec.~\ref{sec2}, we continue to calculate the complex susceptibility for GLE with reset of the memory kernel. The first step is to substitute the Laplace transform of the memory kernel (Eq.~\ref{eq:pl-kernel-laplace-reset}) in $\hat{h}(s) = \frac{1}{s^2 + s \, \hat{\gamma}(s) + \omega^2}$ and calculate the complex susceptibility as defined by Eq.~\ref{eq:complex-susceptibility}. Then, the general form of the susceptibility with resetting takes the form
\begin{align}\label{eq:general-complex-susceptibility}
    \chi(\Omega) = \hat{h}(s=-i\Omega) &= \frac{1}{(-i\Omega)^2 + (-i\Omega) \, \hat{\gamma}_r(-i\Omega) + \omega^2} \nonumber\\&= \frac{1}{-\Omega^2 + (-i\Omega) \frac{-i\Omega+r}{-i\Omega} (-i\Omega+r+b)^{\alpha-1} + \omega^2} \\ \nonumber &= \frac{1}{-\Omega^2 + (-i\Omega+r) (-i\Omega+r+b)^{\alpha-1} + \omega^2}.
\end{align}
For the case without truncation $(b=0)$ and resetting $(r=0)$ we recover the susceptibility reported in \cite{burov2008fractional}
\begin{equation}
    \chi(\Omega) = \hat{h}(-i\Omega) = \frac{1}{-\Omega^2 + (-i\Omega)^{\alpha} + \omega^2},
\end{equation}
also calculated in Eq.~\ref{eq:standard-case-susceptibility} with $b=0$. This was previously obtained in the context of the fractional Klein–Kramers equation in the high-damping regime \cite{barkai2000fractional,coffey2002inertial,coffey2002inertial2,coffey2012langevin}. This form is known as the generalized Rocard equation \cite{rocard1933analyse,scaife1998principles} and reduces to the usual susceptibility of the damped oscillator when $\alpha=1$.

The general form of the response, $R(\Omega)=|\chi(\Omega)|$, with resetting for the case of truncated power-law memory kernel is
\begin{equation} \label{eq:general-response}
    R(\Omega) = \frac{1}{\sqrt{(-\Omega^2+\omega^2+C \cos{(\phi)})^2+(C\sin{(\phi)})^2}} = \frac{1}{\sqrt{S(\Omega)}},
\end{equation}
where
\begin{align}
    \nonumber
    C = \left(\sqrt{r^2+\Omega^2}\right) \left(\sqrt{(r+b)^2+\Omega^2}\right)^{\alpha-1} \quad \text{and} \quad 
    \phi = (1-\alpha) \arctan{\left(\frac{\Omega}{r+b}\right)} - \arctan{\left(\frac{\Omega}{r}\right)}.
\end{align}
The limit of Eq.~\ref{eq:general-response} as $r\to 0$, for the case of a free particle and without truncation ($\omega=b=0$), is
\begin{equation}\label{eq:response-without-reset}
    \lim_{r\to 0} R(\Omega, \omega=0, b=0) = \frac{1}{\sqrt{\Omega^{4} - 2 \Omega^{2+\alpha} \cos{\left(\frac{\pi\alpha}{2}\right)} + \Omega^{2\alpha}}},
\end{equation}
where we have used $\lim_{r\to 0} \arctan{\left(\frac{\Omega}{r} \right)} = \frac{\pi}{2}$ for $\Omega>0$. Eq.~\ref{eq:response-without-reset} corresponds to the standard case without resetting (Eq. 74 with $\gamma=1$ in \cite{burov2008fractional}) and can also be found by taking the limit of Eq.~\ref{eq:response-standard-case} as $b\to 0$. To show the existence of resonance, we look for the solutions of $d R(\Omega)/d\Omega = 0$, for some $\Omega>0$. The response $R(\Omega)$ is given by Eq.~\ref{eq:general-response} and its derivative with respect to $\Omega$ reads
\begin{equation}\label{eq:derivative-R}
    \frac{dR}{d\Omega} = -\frac{1}{2} S^{-\frac{3}{2}}\frac{dS}{d\Omega}.
\end{equation}
Solving $\frac{dR}{d\Omega}=0$ is equivalent to solving $\frac{dS}{d\Omega}=0$. For the standard case ($r=\omega=b=0$), $\frac{dS}{d\Omega}$ is the derivative of the expression inside the square root of Eq.~\ref{eq:response-without-reset}.

We next analyze the response function and stochastic resonance under resetting by considering several specific cases. In particular, we first examine the impact of resetting of the memory kernel on a free GLE particle, then investigate how the potential influences the response, and finally explore the consequences of truncation.

\subsection{Free particle}

\begin{figure}[h!]
\centering
\includegraphics[width=15cm]{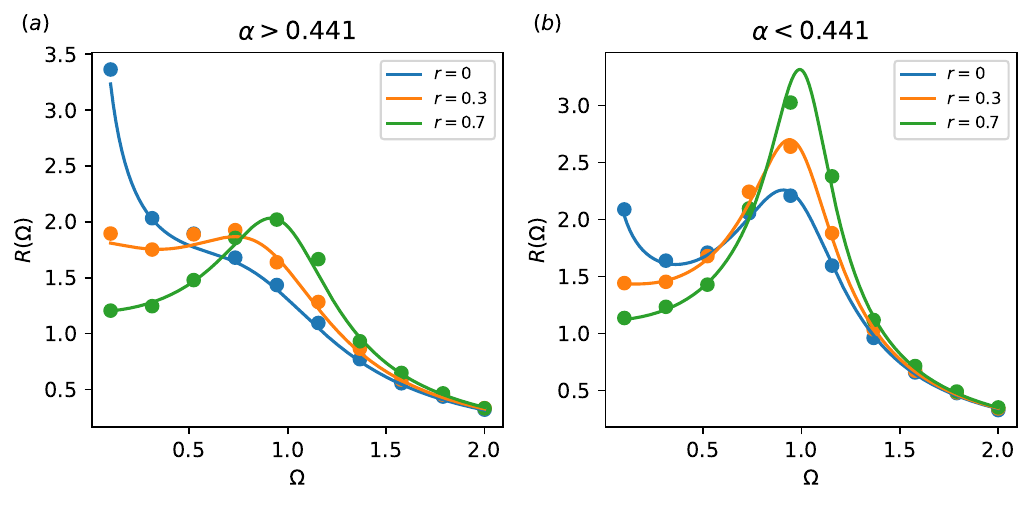}
\caption{The response of GLE to an oscillating field for a free particle under stochastic resetting of the memory kernel without truncation. $(a)$ is for $\alpha=0.5$. $(b)$ is for $\alpha=0.3$. The circles represent results from numerical simulations.}
\label{fig:sr-b0-w0-specific-r}
\end{figure}

To keep things simpler, we first consider the case of an unbounded particle $(\omega=0)$ without truncation $(b=0)$. In this case the response with resetting (Eq.~\ref{eq:general-response}) is
\begin{equation} \label{eq:response-resetting-free}
    R(\Omega) = \frac{1}{\sqrt{\Omega^{4} - 2 \Omega^{2} \left(\Omega^{2} + r^{2}\right)^{\frac{\alpha}{2}} \cos{\left(\alpha \arctan{\left(\frac{\Omega}{r} \right)}\right)} + \left(\Omega^{2} + r^{2}\right)^{\alpha}}} = \frac{1}{\sqrt{S(\Omega)}}.
\end{equation}
As mentioned earlier, the frequency at which the response reaches its maximum is used to identify the resonance phenomenon. To begin exploring the effect of resetting on stochastic resonance, we analyze the behavior of Eq.~\ref{eq:response-resetting-free} for different resetting rates shown in Fig.~\ref{fig:sr-b0-w0-specific-r}. The function $R(\Omega)$ decreases monotonically when $r=0$, but begins to exhibit a distinct local maximum as $r$ increases (Fig.~\ref{fig:sr-b0-w0-specific-r}(a)). In particular, for $r=0.3$ and $r=0.7$, $R(\Omega)$ develops a peak at intermediate values of $\Omega$, reflecting the emergence of a resonance-like behavior in a parameter region in which it was previously absent. Alternatively, we also examine the parameter regime where resonance is observed even when there is no resetting ($r=0$). Once resetting is introduced (Fig.~\ref{fig:sr-b0-w0-specific-r}(b)), for instance at $r=0.3$, the resonance peak grows in magnitude and shifts to slightly higher $\Omega$. As resetting increases further, for example to $r=0.7$, the maximum becomes substantially more pronounced, highlighting the influence of $r$ on both the height of the peak and the frequency at which it appears.

\begin{figure}[h!]
\centering
\includegraphics[width=16cm]{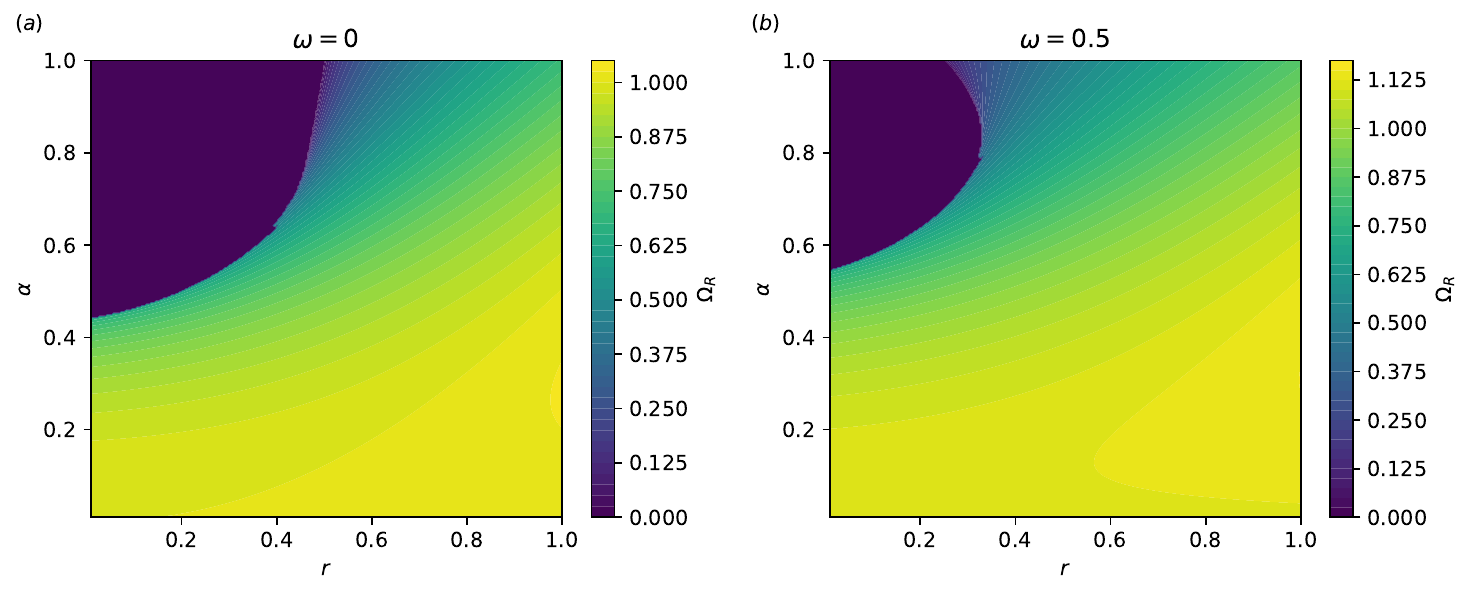}
\caption{Phase diagrams for free (subplot (a)) and harmonically bounded particle (subplot (b)) without truncation ($b=0$). The dark region is a no-resonance phase, whereas the rest is a resonance region. The legend is for the value of the resonant frequency, $\Omega_R$.}
\label{fig:2phase_diagrams}
\end{figure}

To further explore the interplay between the resetting rate and $\alpha$ we now turn our attention to a broader analysis by constructing a phase diagram. This allows us to visualize how the maximum of the response function changes over a wider range of parameters, providing a more comprehensive picture of the resonance behavior. To calculate the maximum of Eq.~\ref{eq:response-resetting-free} we analyze $\frac{dS}{d\Omega}$ (due to Eq.~\ref{eq:derivative-R}) which takes the form
\begin{equation} \label{eq:condition-resonance}
    \frac{dS}{d\Omega} = 4\Omega^3 + 2\alpha\Omega z^{2\alpha-2} - 4\Omega z^{\alpha}\cos{(\alpha \theta)} - 2\alpha\Omega^3 z^{\alpha-2} \cos{(\alpha \theta)} + 2\alpha \Omega^2 z^{\alpha} \sin{(\alpha \theta)} \frac{r}{r^2+\Omega^2},
\end{equation}
where $z=\sqrt{r^2 + \Omega^2}$ and $\theta = \arctan{\left(\frac{\Omega}{r}\right)}$. Specifically, we are looking for the solutions of $\frac{dS}{d\Omega}=0$. Deriving a closed-form solution for this equation is nontrivial, so we instead employ numerical methods. In Fig.~\ref{fig:2phase_diagrams}(a), we present a phase diagram showing how the maximum of the response function varies with respect to the resetting rate $r$ and $\alpha$. At small values of $r$ and sufficiently large $\alpha$, the system remains in the non-resonant regime (upper-left purple region). As $r$ increases (or $\alpha$ decreases), the system transitions to a resonant regime, where the maximum frequency $\Omega$ at which the peak occurs rises from near zero to about 1.

\subsection{Harmonically bounded particle and the effect of truncation}

\begin{figure}[h!]
\centering
\includegraphics[width=15cm]{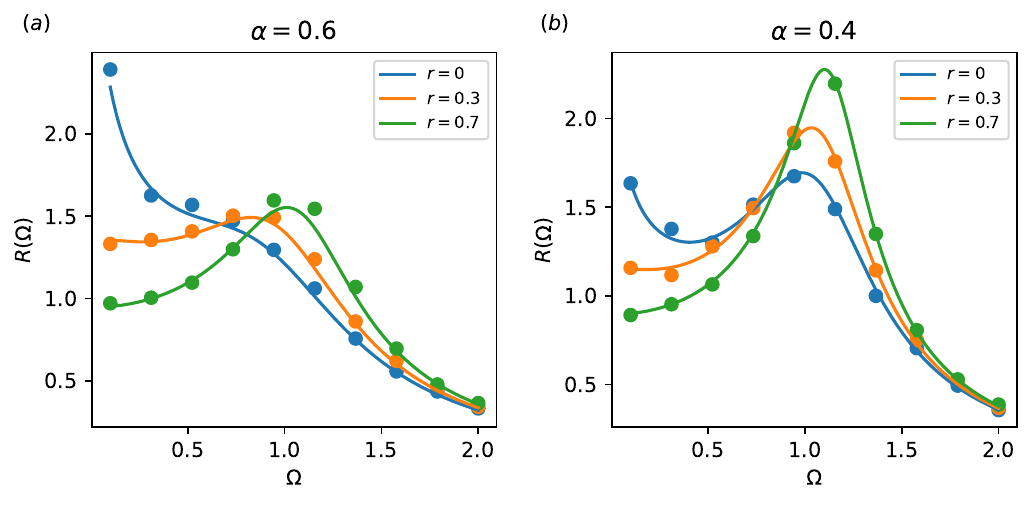}
\caption{The response of GLE to an oscillating field for a harmonically bounded particle $(\omega=0.5)$ under stochastic resetting of the memory kernel without truncation. The circles represent results from numerical simulations.}
\label{fig:sr-b0-w05-specific-r}
\end{figure}

In this section, we begin by analyzing the effect of resetting on the response of the fractional Langevin equation (FLE) when driven by a harmonic force with the potential function $V(x)=\frac{1}{2}\omega^2x^2$. For this case of harmonically bounded particle $(\omega>0)$ and without truncation $(b=0)$ the response with resetting (Eq.~\ref{eq:general-response}) is
\begin{equation}\label{eq:response-resetting-potential}
    R(\Omega) = \frac{1}{\sqrt{(-\Omega^2+\omega^2)^2+2(-\Omega^2+\omega^2)(r^2+\Omega^2)^{\frac{\alpha}{2}}\cos{(\alpha \arctan{(\frac{\Omega}{r})})+(r^2+\Omega^2)^{\alpha}}}} = \frac{1}{\sqrt{S(\Omega)}}.
\end{equation}
We again focus on the resonance points in the response to the applied field, that is, the points at which $R(\Omega)$ is maximized. Note that $R(\Omega)$ now also varies with the potential frequency, $\omega$. To understand the interplay between $\omega$ and $r$, we plot Eq.~\ref{eq:response-resetting-potential} for different resetting rates and $\omega$ in Fig.~\ref{fig:sr-b0-w05-specific-r}. For larger $\alpha$ (e.g., $\alpha=0.6$ in Fig.~\ref{fig:sr-b0-w05-specific-r}(a), we observe a kind of transition from a mostly monotonic decay (the standard case without resetting) to a more resonant-like curve (at some positive resetting rate). On the other hand, for smaller $\alpha$ (e.g., $\alpha=0.4$ in Fig.~\ref{fig:sr-b0-w05-specific-r}(b), a resonant peak is already present even without resetting. Increasing $r$ then shifts the position of this maximum in frequency and enhances the response. To obtain a more detailed picture of the influence of $\omega$ we again construct an $r-\alpha$ phase plane for the solutions of $\frac{dS}{d\Omega}$, shown in Fig.~\ref{fig:2phase_diagrams}(b). The diagram clearly indicates, as expected, that incorporating a harmonic potential contracts the non-resonance region while amplifying the response within the resonance region.

\begin{figure}[h!]
\centering
\includegraphics[width=15cm]{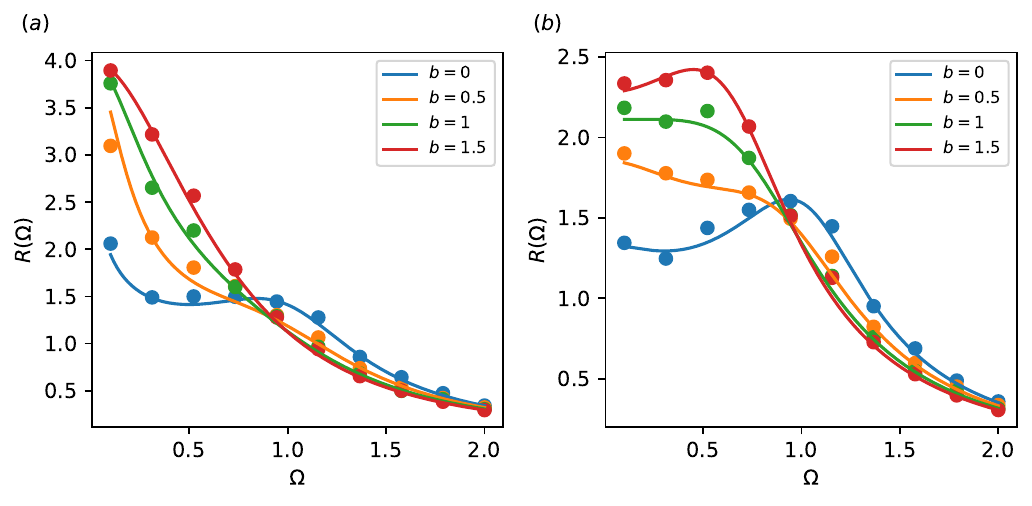}
\caption{The effect of truncation on the response in the case of (a) no resetting, (b) resetting with rate $r=0.25$. Parameters: $\alpha=0.5$ and $\omega=0.5$ for both subplots.}
\label{fig:response-b}
\end{figure}

\begin{figure}[h!]
\centering
\includegraphics[width=17cm]{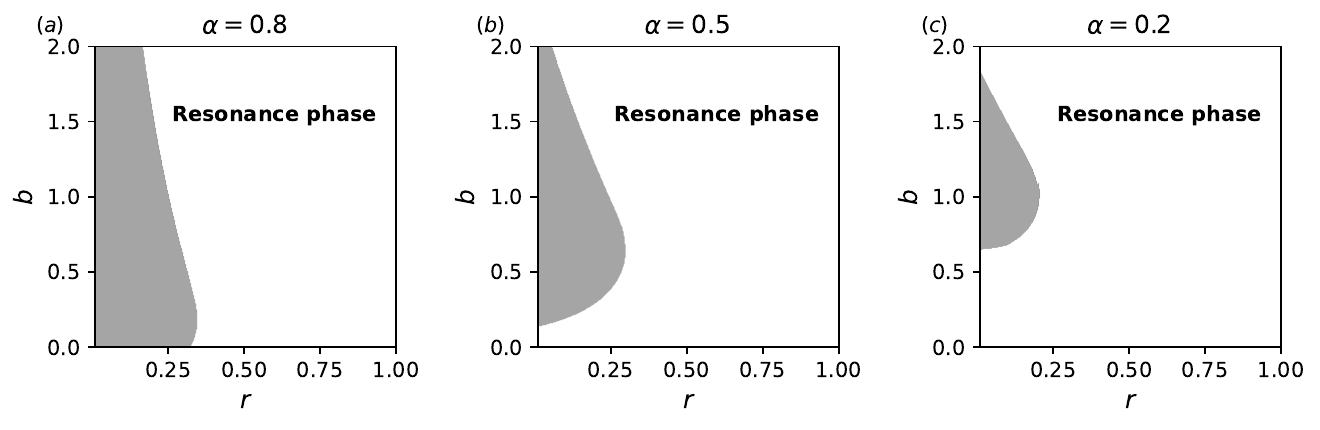}
\caption{$r-b$ phase diagram for a harmonically bounded particle ($\omega=0.5$) for different values of $\alpha$. The gray region is a no-resonance phase, whereas the rest is a resonance region.}
\label{fig:resonance-phase-r-b}
\end{figure}

Finally, we analyze the dependence of the resonant behavior on the truncation parameter $b$. It is known that the resonant peak dissapears as the truncation parameter $b$ becomes larger \cite{liemert2017generalized}. This result is shown in Fig.~\ref{fig:response-b}(a). In order to understand the interrelated effect of $b$, $\omega$ and $\alpha$ coupled with the stochastic resetting mechanism on the resonance phenomenon, we analyze the general form of the response function Eq.~\ref{eq:general-response}. We observe an interesting effect of resetting: for some values of $\alpha$ and $\omega>0$, resonance appears at both low and high values of the truncation parameter $b$, while the resonant peak is absent at intermediate values of $b$ (see Fig.~\ref{fig:response-b}(b) and Fig.~\ref{fig:resonance-phase-r-b}(b,c)). However, this effect disappears for larger $\alpha$ (Fig.~\ref{fig:resonance-phase-r-b}(a)).

\subsection{Dielectric loss and double-peak phenomenon}

In many systems, both the real and imaginary components of susceptibility are measurable quantities, which motivates a detailed investigation of their behavior. The general form of the complex susceptibility, Eq.~\ref{eq:general-complex-susceptibility}, can be written in the following way
\begin{align}
\chi(\Omega)&=\chi^{\prime}(\Omega)+i\,\chi^{\prime \prime}(\Omega) =\\[1mm]
\nonumber \chi^{\prime}(\Omega)&=\frac{\omega^2-\Omega^2+R_1\,R_2^{\,\alpha-1}\cos\Bigl[\theta_1+(\alpha-1)\theta_2\Bigr]}
{\Bigl\{\omega^2-\Omega^2+R_1\,R_2^{\,\alpha-1}\cos\bigl[\theta_1+(\alpha-1)\theta_2\bigr]\Bigr\}^2+\Bigl\{R_1\,R_2^{\,\alpha-1}\sin\bigl[\theta_1+(\alpha-1)\theta_2\bigr]\Bigr\}^2},\\[1mm]
\nonumber \chi^{\prime \prime}(\Omega)&=\frac{R_1\,R_2^{\,\alpha-1}\sin\Bigl[\theta_1+(\alpha-1)\theta_2\Bigr]}
{\Bigl\{\omega^2-\Omega^2+R_1\,R_2^{\,\alpha-1}\cos\bigl[\theta_1+(\alpha-1)\theta_2\bigr]\Bigr\}^2+\Bigl\{R_1\,R_2^{\,\alpha-1}\sin\bigl[\theta_1+(\alpha-1)\theta_2\bigr]\Bigr\}^2},
\end{align}
where
\begin{equation*}
R_1=\sqrt{r^2+\Omega^2},\quad R_2=\sqrt{(r+b)^2+\Omega^2},\quad \theta_1=\arctan\frac{\Omega}{r},\quad \theta_2=\arctan\frac{\Omega}{r+b}.
\end{equation*}
For the case without truncation and resetting, $b = r = 0$, we have $R_{1} = |\Omega|$ and $R_{2} = |\Omega|$ (with the angles $\theta_{1} = \theta_{2} = \tfrac{\pi}{2}$ for $\Omega > 0$). From this, $R_{1}\, R_{2}^{\,\alpha - 1} = |\Omega|^\alpha$ and $\theta_{1} + (\alpha - 1)\,\theta_{2} = \frac{\pi\,\alpha}{2}$, which corresponds to the complex susceptibility in~\cite{burov2008fractional}.

\begin{figure}[h]
\centering
\includegraphics[width=15cm]{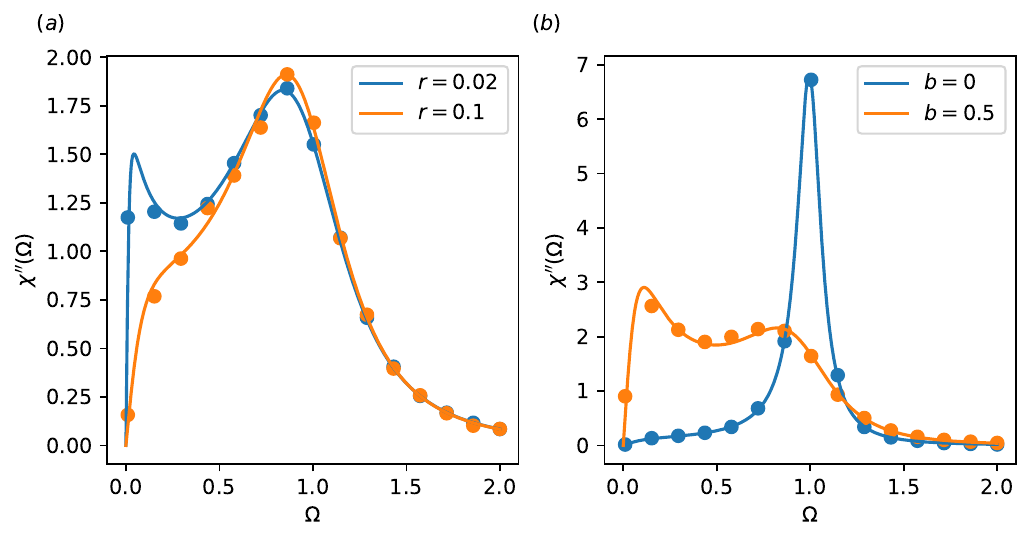}
\caption{Double peak phenomenon of the imaginary part of complex susceptibility under resetting. (a) $\alpha=0.4$ without truncation, (b) $\alpha=0.1$ with resetting rate $r=0.1$.}
\label{fig:double-peak-simulations}
\end{figure}

\begin{figure}[h]
\centering
\includegraphics[width=17cm]{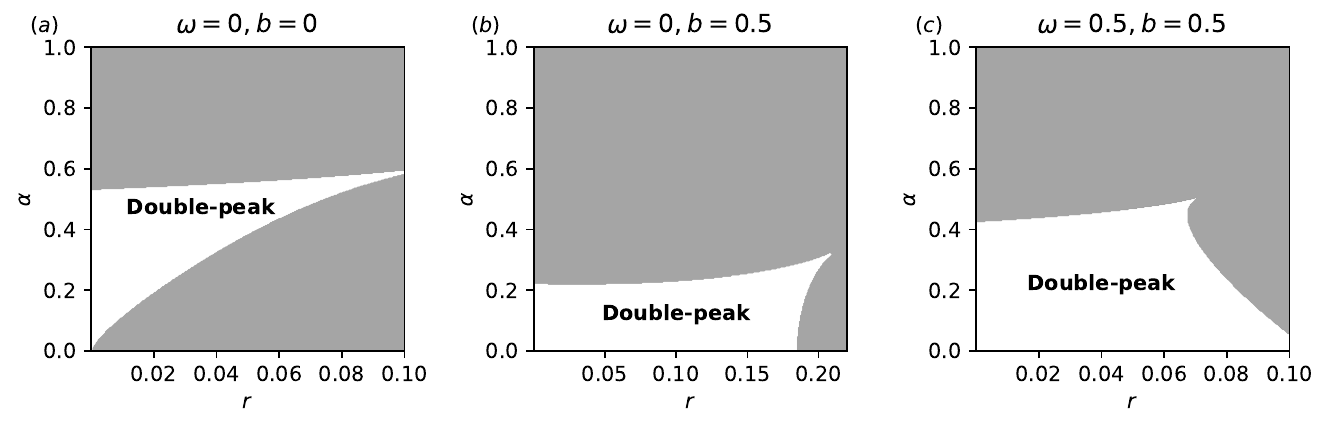}
\caption{Phase diagram of the imaginary part of complex susceptibility under resetting. The gray region is one-peak phase, whereas the white region is double-peak phase.}
\label{fig:double-peak-phase}
\end{figure}

In this subsection we explore the behavior of the imaginary part $\chi^{\prime \prime}(\Omega)$ which is commonly referred to as "the loss". An interesting phenomenon that is observed in the behavior of $\chi^{\prime \prime}(\Omega)$ is the double-peak phase that occurs for some $\alpha^{\prime}$s due to the fractional dynamics. Such phenomena have been observed in relaxation processes of super-cooled liquids \cite{gotze1992relaxation}. Burov and Barkai~\cite{burov2008fractional} calculated critical thresholds for the parameter $\alpha$, which indicate whether "the loss" exhibits a single peak or a double peak, depending also on the angular frequency $\omega$ of the harmonic potential. In particular, for a free particle, the critical threshold for transitioning from a double-peak to a single-peak phase is $\alpha \approx 0.527$. Here, we recover the same result numerically from the general form of the imaginary part of the susceptibility. From Fig.~\ref{fig:double-peak-simulations}(a) we observe that for certain value of $\alpha$ and no truncation, a small amount of resetting can preserve the double-peak phenomenon; however, increasing the resetting rate further leads to a transition from a double-peak to a single-peak phase. Moreover, we observe an interesting interplay between truncation and resetting (Fig.~\ref{fig:double-peak-simulations}(b)). Specifically, for some truncation parameter $b$, a transition from one-peak to double-peak phase emerges even at $r=0.1$. However, the loss returns to a single peak for $r$ greater than 0.17 for $b=0.5$, as shown in the phase diagram in Fig.~\ref{fig:double-peak-phase}. This diagram indicates that without truncation, increasing the resetting rate causes the loss to transition from a double-peak regime to a single-peaked one (Fig.~\ref{fig:double-peak-phase}(a)). However, the effect of truncation recovers the double-peak structure for some resetting rates (Fig.~\ref{fig:double-peak-phase}(b)).

\begin{figure}[h!]
\centering
\includegraphics[width=12cm]{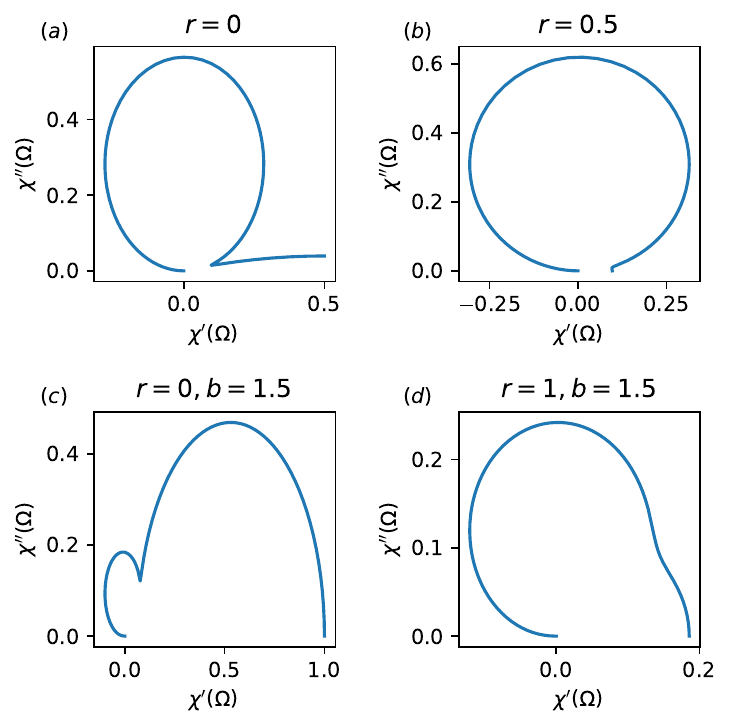}
\caption{Cole-Cole diagrams of complex susceptibility. Parameters: $\alpha=0.1, \gamma=10, \omega=1$. The first row is the case without truncation, $b=0$.}
\label{fig:cole-cole1}
\end{figure}

The observed double-peak phenomenon appears due to behavior of the friction term like an elastic force for small $\alpha$, thereby inducing oscillatory dynamics within the system. This interpretation is explained by the Cole-Cole diagrams of complex susceptibility (Fig.~\ref{fig:cole-cole1}). Specifically, for small $\alpha=0.1$ (Fig.~\ref{fig:cole-cole1}(a)) two distinct types of normal susceptibility coexist: a Debye type for normal damped oscillator (left side of the plot) \cite{kubo2012statistical} and in the right side of the plot a monotonic relaxation (Van-Vleck-Weisskopf-Fröhlich type, \cite{kubo2012statistical}). This leads to the conclusion that the system exhibits two characteristic frequencies: the lower frequency governs the monotonic decay, while the higher frequency is associated with oscillatory behavior \cite{burov2008fractional}. When resetting is introduced (Fig.~\ref{fig:cole-cole1}(b)), the monotonic decay component vanishes, leaving only oscillatory relaxation. On the other hand, the truncation has a similar effect to higher $\alpha$ such that we have some mixed behavior as shown in (Fig.~\ref{fig:cole-cole1}(c)). Furthermore, the effect of truncation is overcome by resetting (Fig.~\ref{fig:cole-cole1}(d)), leading again to an oscillatory relaxation akin to Van Vleck–Weisskopf–Fröhlich type.

\section{Simulation methodology}\label{sec:simulation-methodology}

In this section we outline the procedure for the numerical solution of the GLE system used to confirm the analytical results in previous sections. Unlike the simpler, memory-free Langevin equation, simulating the generalized Langevin equation is not a trivial task. As noted in \cite{biswas2025resetting}, Eq.~\ref{eq:gle-free} is difficult to solve numerically due to its non-Markovian nature, because, first, the convolution term requires storing the entire velocity history and computing convolutions at each measurement step, which becomes computationally expensive and second, generating the correlated noise $\zeta(t)$ for long time spans also demands large data storage and intensive computations, especially when many trajectories are needed for accurate statistical averaging. Consequently, both the memory and processing requirements grow rapidly, making straightforward numerical simulations impractical. To overcome these numerical challenges, we utilize the Markovian embedding method explained in more detail in the next subsection. After successfully developing a simulation scheme for the GLE under reset of the memory kernel, the second part of this section presents a numerical method to estimate the system’s frequency response. An implementation of the simulation methodology is available at the following \href{https://github.com/pero-jolak/generalized-langevin-equation}{link}.

\subsection{GLE with resetting: A Markovian embedding}

As explained in \cite{goychuk2012viscoelastic} in detail, the method of Markovian embedding represents the complex, non-Markovian system as a projection of a higher-dimensional Markov process onto a reduced subspace of relevant variables. By elastically coupling a central Brownian particle to a set of auxiliary Brownian particles—each subjected to purely viscous friction and white thermal Gaussian noise—the approach effectively simulates a broad range of viscoelastic environments. Remarkably, the embedding requires only a small number of auxiliary particles, enabling accurate modeling of subdiffusive behavior over many temporal decades while maintaining computational efficiency. Specifically, the fluctuation-dissipation relation is obeyed by expanding the kernel into a sum of exponentials as in Eq.~\ref{eq:sum-exponentials} and the noise $\zeta(t)$ into a sum of Ornstein-Uhlenbeck processes with corresponding exponentially decaying autocorrelation functions.

We begin with the general truncated power-law memory kernel under resetting as defined by Eq.~\ref{eq:kernel-resetting} and define it as a target kernel, $K_{\mathrm{target}}(t)$. The Markovian embedding method exactly recovers the original GLE only when the memory kernel represents the exponential decay \cite{wisniewski2024dynamics,kupferman2004fractional}. However, in our case, the power-law memory kernel can be approximated by a finite sum of exponentials\footnote{This approach is similar to the method of determining the coefficients in a Prony series in viscoelastic materials \cite{tzikang2000determining}.}
\begin{equation} \label{eq:sum-exponentials}
  K_{\mathrm{model}}(t) = \sum_{i=1}^N c_i\, e^{-\,t / \tau_i}.
\end{equation}
Given a set of fitting points \(\{t_j\}\), we form the design matrix $A_{i j} = \exp\Bigl(-\frac{t_i}{\tau_j}\Bigr)$ and let
\begin{equation*}
  \mathbf{c}
  =
  (c_1, c_2, \dots, c_N)^T,
  \qquad
  \mathbf{K}_{\mathrm{target}}
  =
  \bigl(K_{\mathrm{target}}(t_1),\dots,K_{\mathrm{target}}(t_M)\bigr)^T.
\end{equation*}
We evaluate the kernel at \(M\) points, \(\{t_1, t_2, \dots, t_{M}\}\), logarithmically spaced over the range $t \in [dt, T]$ to ensure that the kernel is well-sampled across different time scales. We then solve the constrained least-squares problem
\begin{equation}
  \min
  \,\bigl\|\,
    A\,\mathbf{c}
    -
    \mathbf{K}_{\mathrm{target}}
  \bigr\|^2,
  \quad\text{subject to }
  c_i \ge 0.
\end{equation}

where $A\,\mathbf{c}$ is $K_{\mathrm{model}}$. Numerically, this is done with the L-BFGS-B algorithm~\cite{byrd1995limited,zhu1997algorithm}.

\begin{figure}[h!]
\centering
\includegraphics[width=15cm]{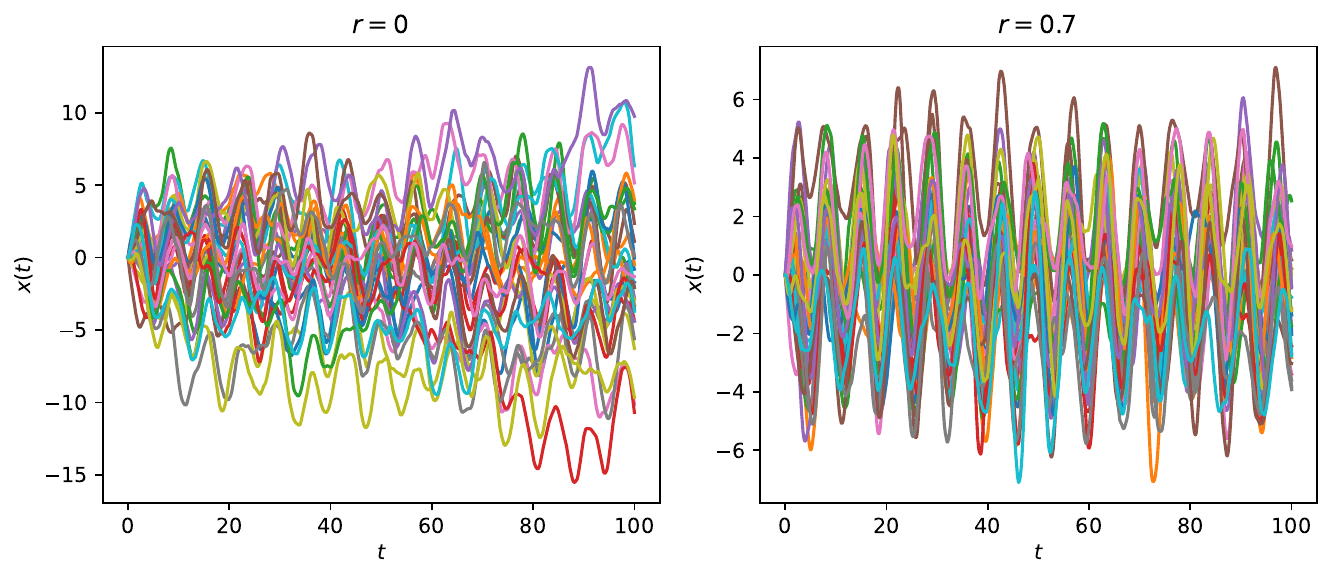}
\caption{Typical trajectories of the process for $\alpha=0.5$. Free particle without truncation ($\omega=b=0$) and $\Omega\approx 0.915$ which is the resonant frequency when $r=0.7$. Left: without resetting. Right: with resetting rate $r=0.7$.}
\label{fig:trajs}
\end{figure}

After finding the coefficients \(\{c_i\}\), we simulate a system of stochastic differential equations, which corresponds to a Markovian embedding of the GLE (Eq.~\ref{eq:gle-free})~\cite{goychuk2009viscoelastic, goychuk2012viscoelastic,wisniewski2024dynamics,siegle2010markovian,biswas2025resetting}:
\begin{equation}
\left\{
\begin{aligned}
\frac{\mathrm{d}x}{\mathrm{d}t} &= v, \\
\frac{\mathrm{d}v}{\mathrm{d}t} &= \sum_{i=1}^N z_i - \omega^2 x + A_0 \cos(\Omega t), \\
\frac{\mathrm{d}z_i}{\mathrm{d}t} &= -\frac{z_i}{\tau_i} - c_i v 
      + \sqrt{\frac{2c_i}{\tau_i}}\, \xi_i(t), \quad i=1,\dots,N,
\end{aligned}
\right.
\end{equation}
where $N$ is the number of exponentials used to approximate the kernel and each \(\xi_i(t)\) is an independent Gaussian white noise process with zero mean and autocorrelation
\(\langle \xi_i(t)\,\xi_j(t^{\prime}) \rangle = 2\,\delta_{ij}\,\delta(t - t^{\prime})\). The particle's position is initialized at $x(0)=0$. Its initial velocity $v(0)$ is drawn from a standard normal distribution, and each $z_i(0)$ is drawn from a normal distribution with mean 0 and standard deviation $\sqrt{c_i}$. As an illustration, typical trajectories of the process are shown graphically in Fig.~\ref{fig:trajs}. The parameters are chosen such that the trajectories in the left figure are in non-resonance regime, while those in the right figure are in a regime where there is a resonant peak for some frequency. Interestingly, for a particular resetting rate (as we enter from a resonant to a non-resonant regime) the trajectories are in 'sync' which is a qualitative indication of resonance.

\subsection{Estimating the frequency response}

Previously, we showed how to simulate the GLE trajectories under resetting. We use these $N$ trajectories\footnote{This $N$ should not be confused with the number of exponentials in the previous subsection.}, each described by a discrete time series
\begin{equation*}
x_k(j) = x_k\bigl(t_j\bigr), 
\quad t_j = j\,\Delta t, 
\quad j = 0,1,\ldots,M-1, 
\quad k = 1,2,\ldots,N,
\end{equation*}
with sampling interval \(\Delta t\). The total duration of the time interval is $T = M \,\Delta t$. For each trajectory \(k\), the Discrete Fourier Transform (DFT) \cite{cooley1965algorithm} of \(x_k(j)\) is
\begin{equation}
\widehat{X}_k(\Omega_m)
=
\sum_{j=0}^{M-1} x_k(j)\, e^{-\,i \,2\pi\,\frac{m\,j}{M}},
\end{equation}
where
\begin{equation*}
\Omega_m 
=
\frac{2\pi m}{M\,\Delta t},
\quad m = 0,1,\ldots,M-1.
\end{equation*}
The reference signal (external forcing function) is $F(t) = \cos\bigl(\Omega\, t\bigr)$, which is sampled at the same times \(t_j\): 
$F(j) = F\bigl(t_j\bigr) 
      = \cos\bigl(\Omega\,j\,\Delta t\bigr)$. Its DFT is
\begin{equation}
\widehat{R}(\Omega_m)
=
\sum_{j=0}^{M-1} F(t_j)\, e^{-\,i\,2\pi\,\frac{m\,j}{M}}.
\end{equation}
Then, the following ratio represents the complex susceptibility (Eq.~\ref{eq:complex-susceptibility}) estimated numerically
\begin{equation}
\chi_k(\Omega_m) =
\frac{\widehat{X}_k(\Omega_m)}{\widehat{R}(\Omega_m)},
\end{equation}
which can be regarded as normalizing the transform of \(x_k\) by the transform of the periodic force. Next, we define the discrete frequencies (in cycles/second),
$f_m =
\frac{\Omega_m}{2\pi} =
\frac{m}{M \,\Delta t}$, and find the index \(m^*\) such that \(f_{m^*}\) is closest to the driving frequency used to simulate the trajectories, \(f_{\text{drive}}=\Omega/2\pi\):
\begin{equation}
m^* 
=
\arg\min_{m} \;\bigl|\,f_m - f_{\text{drive}}\bigr|.
\end{equation}

At the index \(m^*\), we compute the amplitude, $ A_k  =
\bigl|\chi_k(\Omega_{m^*})\bigr|$ and the imaginary part, $I_k = \Im\bigl(\chi_k(\Omega_{m^*})\bigr)$. We do this for each trajectory \(k = 1,2,\ldots,N\) and finally define the average amplitude and the average imaginary response over all \(N\) trajectories
\begin{equation}
\text{Mean Amplitude} 
=
\frac{1}{N}\sum_{k=1}^N A_k,
\quad
\text{Mean Imaginary Response (Loss)}
=
\frac{1}{N}\sum_{k=1}^N I_k.
\end{equation}

\section{Summary}\label{sec:discussion}

We examine how the mechanism of stochastic resetting of the memory kernel influences the response function and the loss in case of a generalized Langevin equation with truncated power-law memory kernel and an external time dependent field. By systematically varying the system parameters, our study reveals that stochastic resetting can effectively trigger resonance in regimes where standard cases of GLE would predict its absence. Specifically, for a free particle, a resonant behavior occurs when the kernel exponent $\alpha$ is below a certain threshold $\alpha_R$, however, with resetting, a resonant peak can emerge even when $\alpha$ exceeds $\alpha_R$. Moreover, as is known, when the truncation parameter $b>0$ increases, the resonant peak tends to vanish. In this case, the resetting mechanism interacts with the angular frequency of the potential in such a way that resonance can be induced even with stronger truncation. These observations stem from the fact that the resetting mechanism modifies the noise's correlation structure via the memory kernel to comply with the fluctuation-dissipation theorem, thereby inducing the phenomenon of stochastic resonance. On the other hand, for a fixed $\alpha$ and without truncation, only a small enough resetting rate can maintain the double-peak behavior in the imaginary part of the complex susceptibility. Additionally, higher resetting rates drive the system from a double-peak to a single-peak regime, while the introduction of truncation can recover the double-peak phenomenon at some resetting rates. In terms of the Cole-Cole plot of the complex susceptibility, resetting of the memory kernel leads to an oscillatory relaxation similar to a Van Vleck–Weisskopf–Fröhlich type. Finally, we validate our analytical findings with numerical simulations of the GLE under reset of the memory kernel, utilizing a Markovian embedding approach.

The insights gained here not only enhance our theoretical understanding of non-equilibrium systems with memory, but also pave the way for potential experimental realizations in complex systems where tuning the resetting rate could lead to controlled resonance phenomena.

\section*{Acknowledgments}

PJ, LK and TS acknowledge the financial support from the German Science Foundation (DFG, Grant number ME~1535/12-1) and the Alliance of International Science Organizations (Project No.~ANSO-CR-PP-2022-05). TS is also supported by the Alexander von Humboldt Foundation.

\bibliography{bibliography}
\bibliographystyle{unsrt}

\end{document}